\documentclass[]{raa} 

\usepackage{graphicx,times}             
\usepackage{aas_macros} 
\usepackage{natbib}

\bibpunct{(}{)}{;}{a}{}{,}
\usepackage[colorlinks,linkcolor=blue,urlcolor=blue,citecolor=blue]{hyperref}

\usepackage{amssymb}
\usepackage{savesym}
\usepackage{amsmath}
\savesymbol{iint}
\usepackage{txfonts}
\restoresymbol{TXF}{iint}
\usepackage{booktabs}
\usepackage{amsmath}
\usepackage{epsfig}
\usepackage{mathtools} 	

\usepackage{graphicx,times}
\usepackage{natbib,amssymb}
\usepackage{epsfig}
\usepackage{float}
\usepackage[utf8]{inputenc}
\usepackage{natbib}
\usepackage{mathrsfs}
\usepackage{ulem}

\graphicspath{{./}{figures/}}

\begin{document}
\title{The Binarity of Early-type Stars from LAMOST Medium-resolution Spectroscopic Survey}

 \volnopage{ {\bf 2018} Vol.\ {\bf X} No. {\bf XX}, 000--000}
 \setcounter{page}{1}

\author{Yanjun~Guo\inst{1,2}
        \and Jiao~Li\inst{3}
        \and Jianping~Xiong\inst{3}         
        \and Jiangdan~Li\inst{1,2} 
        \and Luqian~Wang\inst{1} 
        \and Heran~Xiong\inst{4}
        \and Feng~Luo\inst{3}
        \and Yonghui~Hou\inst{5,6} 
        \and Chao~Liu\inst{2,3} 
        \and Zhanwen~Han\inst{1,2,7}                     
        \and Xuefei~Chen\inst{1,2,7}
        }
\institute{Yunnan observatories, Chinese Academy of Sciences, 
           P.O. Box 110, Kunming, 650011, China\\
           \email{liuchao@bao.ac.cn;cxf@ynao.ac.cn}
           \and
           School of Astronomy and Space Science, University of Chinese Academy of Sciences, Beijing, 100049, People's Republic of China
           \and
           Key Laboratory of Space Astronomy and Technology, National Astronomical Observatories, Chinese Academy of Sciences, Beijing 100101, People's Republic of China
           \and
           Research School of Astronomy and Astrophysics, Mount Stromlo Observatory, The Australian National University, ACT 2611, Australia
           \and
           Nanjing Institute of Astronomical Optics, $\&$ Technology, National Astronomical Observatories, Chinese Academy of Sciences, Nanjing 210042, China
           \and
           School of Astronomy and Space Science, University of Chinese Academy of Sciences
           \and
           Center for Astronomical Mega-Science, Chinese Academy of Sciences, Beijing 100012, China
          }

\abstract
{
Massive binaries play significant roles in many fields. Identification of massive stars, particularly massive binaries, is of great importance. In this paper, by adopting the technique of measuring the equivalent widths of several spectral lines, we identified 9,382 early-type stars from LAMOST medium-resolution survey and divided the sample into four groups, T1 ($\sim$O-B4), T2 ($\sim$B5), T3 ($\sim$B7), and T4 ($\sim$B8-A). The relative radial velocities $RV_{\rm rel}$ were calculated using the Maximum Likelihood Estimation. The stars with significant changes of $RV_{\rm rel}$ and at least larger than 15.57km s$^{-1}$ were identified as spectroscopic binaries. We found that the observed spectroscopic binary fractions for the four groups are $24.6\%\pm0.5\%$, $20.8\%\pm0.6\%$, $13.7\%\pm0.3\%$, and $7.4\%\pm0.3\%$, respectively. Assuming that orbital period ($P$) and mass ratio ($q$) have intrinsic distributions as $f(P) \propto P^\pi$ (1\textless$P$\textless1000 days) and $f(q) \propto q^\kappa$ (0.1\textless$q$\textless1), respectively, we conducted a series of Monte-Carlo simulations to correct observational biases for estimating the intrinsic multiplicity properties. The results show that the intrinsic binary fractions for the four groups are 68$\%\pm8\%$, 52$\%\pm3\%$, 44$\%\pm6\%$, and 44$\%\pm6\%$, respectively. The best estimated values for $\pi$ are -1$\pm0.1$, -1.1$\pm0.05$, -1.1$\pm0.1$, and -0.6$\pm0.05$, respectively. The $\kappa$ cannot be constrained for groups T1 and T2 and is -2.4$\pm0.3$ for group T3 and -1.6$\pm0.3$ for group T4. We confirmed the relationship of a decreasing trend in binary fractions towards late-type stars. No correlation between the spectral type and the orbital period distribution has been found yet, possibly due to the limitation of observational cadence.
}

\titlerunning{The Binarity of Early-type Stars of LAMOST}

\authorrunning{Yanjun Guo et al.}

\maketitle

\section{Introduction}\label{sec:intro}
Massive binaries have long been a topic of great interest in a wide field of astronomy. 
The evolution of binary systems significantly differs from that of single stars \citep{2009Mason,2020Hanzhanwen}. \cite{2012Sana1} have shown that over 71\% of O-type stars interact with their companions. At the end of their evolution, such massive binaries can lead to the formation of double compact objects such as double black holes (DBHs), double neutron stars (DNSs), and neutron star-black hole (NS-BH). In addition, the evolution of massive binary stars is the major channel to form potential gravitational sources \citep{2002Pfahl,2016GW1,2016GW2,2017Tauris,2018XueFei,2020GWbinary}. 

Binary population synthesis (BPS) is a popular method to study the statistical properties of a type of star (birthrate, local space density, etc.) \citep{2020Hanzhanwen}. The multiplicity properties, including the intrinsic binary fraction, distributions of orbital period and mass ratio, are the basic physical inputs for binary population synthesis. Until now, the binary fraction is still uncertain. Varying with spectral type, the binary fraction of stars could be as low as 20\% for late-type stars, reaching up to 80\% for early-type stars \citep{2013Duchene,2017MoeStefano}. For binaries with longer orbital periods, the stars have weak gravitational interaction with each other, resulting in difficulty detecting such binary systems. The orbital period is an essential parameter that can directly relate to dynamical evolution \citep{2013Duchene}. The mass ratio can be used to describe the relationship between the primary star and its companion star.  
The orbital period and the mass ratio are essential parameters to determine whether the binary system would evolve under stable or non-stable mass transfer scenario or through a common envelope stage \citep{1998Kalogera,2010MaizApellaniz}. 

For several decades, large works have been done to investigate the properties of spectroscopic binaries  \citep{1980Garmany,1990Abt,2007Kobulnicky,2012Chini,2012Sana1,2013Sana,2014Sota,2015Dunstall,2015Aldoretta,2016Maiz,2017Almeida}. \cite{1979Abt} have studied the relationship between binary fraction in the field and clusters. \cite{1999Abt} investigated the relationship between age and binary fractions in five open clusters. The relationship between spectral types and binary stars have been studied by \cite{1935Kuiper} and \cite{2010Raghavan}.
However, the knowledge of the multiplicity properties for early-type stars is still limited \citep{2013Duchene,2017MoeStefano,2017Sana}. 
Results are varying from ones adopting different techniques or from the observation biases of different datasets.
For example, spectroscopy is not sensitive enough to detect long period binaries, and speckle interferometry is limited to detect binary systems with a smaller mass ratio \citep{2011Sana}. 

The statistical approach is widely used to investigate the intrinsic multiplicity properties of massive stars. \cite{2007Kobulnicky} developed the two-sided K-S test and a Monte-Carlo approach to correct the observational bias of early-type stars. However, the result reported from \cite{2007Kobulnicky} is highly debated since they estimated the intrinsic binary fraction reaches about 80-100\%, this conclusion seems too high, and the sample is too small to give a reliable conclusion \citep{2003Harries,2005Hilditch,2006Pinsonneault,2006Lucy}. In order to make the statistic simulations match the observations, \cite{2013Sana} improves this method and gives the intrinsic binary fraction to be 51$\%\pm4\%$.

Thanks to the advanced modern telescopic instruments, many optical sky surveys such as the Galactic O-Star Spectroscopic Survey (GOSSS), a high-resolution monitoring program of Southern Galactic O- and WN-type stars (OWN), a high-resolution spectroscopic database of Galactic Northern OB-type stars (IACOB), provide good opportunities to search for massive binary systems \citep{2010MaizApellaniz,2010Barba,2011SimonDiaz,2013MaizApellaniz,2017Sana}. 

The Large Sky Area Multi-Object Fiber Spectroscopic Telescope (LAMOST) is capable of searching at most 4000 objects simultaneously \citep{2012CuiXiangQun,2012ZhaoGang}. It has obtained more than 10 million low-resolution (R=1800) spectra, including about 9 million stellar spectra, during its first stage (LAMOST I) from 2011 to 2018. LAMOST II started in 2018, and stage II includes both low- and medium-resolution (R=7500) spectroscopic surveys (hereinafter LRS and MRS) \citep{2012DengLiCai,2020LiuChao}. In particular, it includes time-domain observations in MRS. 
In this work, we made usage of more than 5 million stellar spectra archived from the LAMOST-MRS database, and the observations were taken between 2018 September and 2019 June.
Moreover, only a few works have been done on early-type stars, and such analyses are performed based on LAMOST low-resolution spectra. Examples include classifying the OB candidates from LAMOST DR5 \citep{2019LiuZhicun}, estimating of absolute magnitude, distance, and binarity of OB stars from LAMOST DR5 \citep{2020Xiangmaosheng}, etc. The purpose of this study is to investigate the binary fraction of early-type stars using the medium resolution spectra from the LAMOST.

In Sec.~\ref{sec:data} we introduce the LAMOST-MRS data. In Sec.~\ref{sec:OBA} we describe our work of analyzing the equivalent widths of several spectral lines to divided the sample into four groups based on their spectral types.
In Sec.~\ref{sec:RV}, we describe the method of calculating the radial velocity and identify spectroscopic binaries. In Sec.~\ref{sec:MCMC all} we report the work of using a Monte-Carlo method to estimate the intrinsic multiplicity parameters. Our summary is in Sec.~\ref{sec:conclusions}.

\section{Data} \label{sec:data}
The LAMOST is a 4-meter Schmidt telescope \citep{2012CuiXiangQun,2012ZhaoGang,2012LuoALi}. It has great potential to search for millions of objects in the northern sky. Many results have been obtained from LAMOST data, especially in Milky Way science, and stellar astrophysics \citep[e.g.][]{2010WuXueBing,2014LiuChao,2015LiJiao,2015LiuChao,2015Zhangbo,2016HouWen,2017LiuChao,2018Tian,2019LiuJifeng,2019LiuZhicun,2020ApJSslam,2020RAAcod,2020Tian}. 

In this work, we use the optical spectra of LAMOST-MRS with the wavelength range of 4950-5350\ \AA\ for the blue band and 6300-6800\ \AA\ for the red band. The wavelength ranges covered by LAMOST MRS include the Mgb series, H$\alpha$ at around 6564 \AA, HeI at around 6679 \AA\ (See Table~\ref{Tab:1}), which is what we want for studying early-type stars. The wavelength calibration is accomplished based on the Sc and Th$-$Ar lamps for MRS spectra. After extracting arc lamp spectra, the Legendre polynomial as a function will be used to fit the measured centroids of the arc lines to describe the relationship between wavelengths and pixels \citep{2021Renjuanjuanwvcalu}, and the typical accuracy of wavelength calibration is 0.05~\AA\ $\rm pixel^{-1}$ for LAMOST-MRS. For late-type stars with medium-resolution spectra, the precision of the radial velocities is around $1 km s^{-1}$ \citep{2020LiuChao}.

\section{Identification of OBA Star}\label{sec:OBA}
\subsection{The Initial Sample}\label{sec:sample set}
We collected a sample of over 800,000 stars from LAMOST-MRS. In these objects, $\sim$500,000 have {\tt{stellar parameters}}\footnote{\url{http://dr7.lamost.org/v1.3/catalogue}} derived from LAMOST stellar parameter pipeline (LASP) \citep[][Luo et al. 2021 in prep]{2015luoaliLRSstellarpara}. The LASP adopts computational algorithms of the Correlation Function Initial (CFI) \citep{2012Dubing} and the Universite de Lyon Spectroscopic analysis Software (ULYSS) \citep{2009KolevauLYSSF}, and obtains $T_\mathrm{eff}$ by minimizing the $\chi^2$ between model spectra and observed spectra \citep{2015luoaliLRSstellarpara}. Due to the limitation of the pipeline, only stellar parameters for A and later type stars i.e. with $T_\mathrm{eff}$ in range of 3100 $-$ 8500 K, are given \citep{2011wuyueLASP}. The typical uncertainty of $T_\mathrm{eff}$ for LRS from the LASP is 110 K \citep{2014wuyueLAMOSTT,2015GaohuaLRSerr}, whilst that for MRS was not given yet. \cite{2020ApJSslam} have shown that the accuracy of simulated MRS spectra is similar to that of LRS for the stars with solar abundances.

Since we were only concerned with the multiplicity of early-type stars in this paper, we then eliminated stars with spectral type later than A5 i.e with estimated $T_\mathrm{eff}$ $<$ 8000 K given by the LASP. We included $\sim$ 300,000 stars without estimated $T_\mathrm{eff}$ in our initial sample because many early-type stars are in this part. However, a broad range of stars of various spectral types is mixed in such a sample. We thus eliminated spurious late-type stars in our sample by the following selection criteria. 

\subsection{Removing stars later than A5}\label{sec:oba cer}
\begin{figure*}
        \centering
	    \includegraphics[scale=0.5]{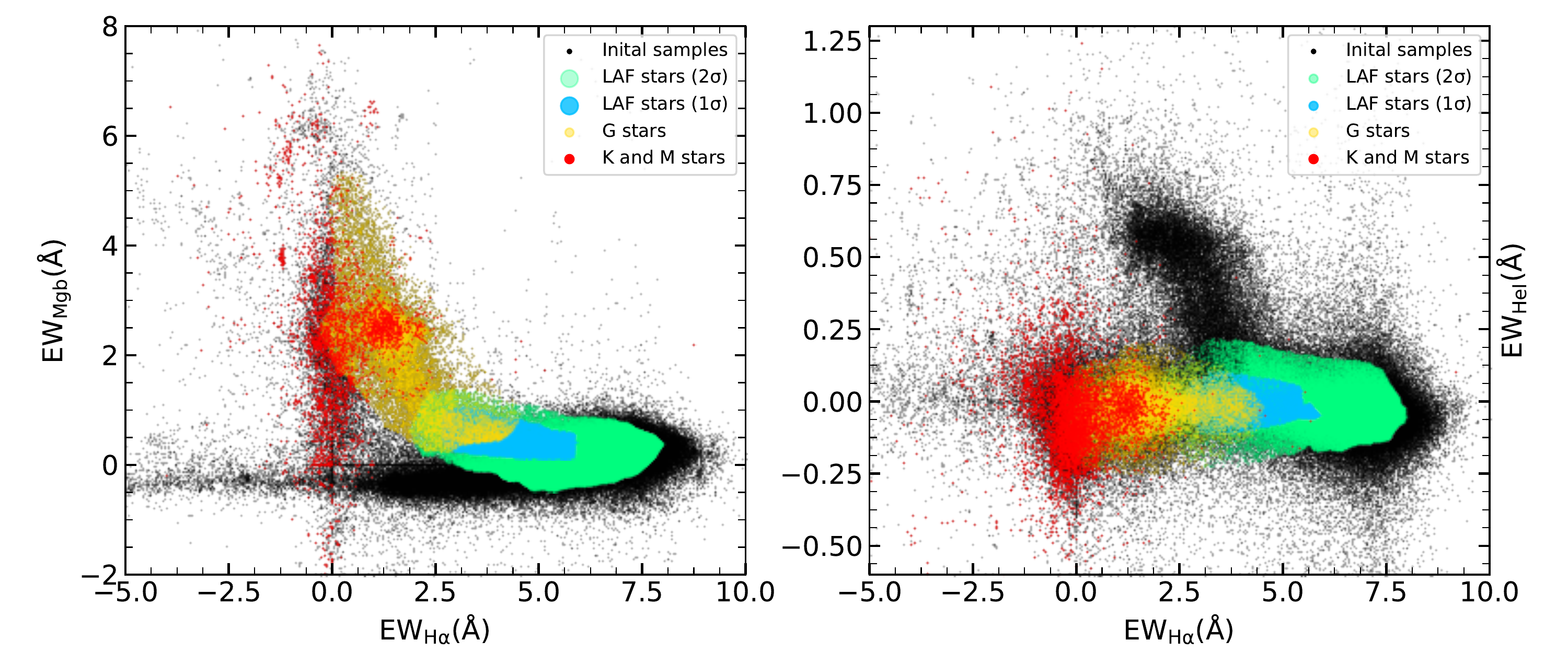}
		\caption{The black dots are our initial sample. The blue and green dots represent the late A- and F-type (LAF) stars in the 1$\sigma$ and 2$\sigma$ contour region of Fig.~\ref{fig:F sigma}, respectively. The yellow dots represent the G-type stars, and the red dots represent M- and K- type stars (for details, see Sec.~\ref{sec:oba cer}).}\label{fig:All candidate}
\end{figure*}

The classic method to classify stars with MK types is comparing their spectra manually with standard stars \citep{2009Gray,2014Gray}. This approach is hard to handle with huge samples of observations. 
The photometric color indices are also widely used in classifying stars, 
but such a method is heavily affected by interstellar reddening. 
It is difficult to calibrate the reddening in the deep sky \citep{2015LiuChao}. 
In this work, we thus employed the equivalent widths (EWs) of several lines to exclude late-type stars (later than A5) in our sample \citep{2015LiuChao}. The EW for a given spectral line is defined as
\begin{equation}
\centerline{ $EW= \int_{}^{}(1-\frac{F_{\rm line}(\lambda)}{F_{\rm con}(\lambda)})d\lambda$,}\label{eq:1}
\end{equation}
where $F_{\rm line}(\lambda)$ is the flux of the spectral line at a given wavelength $\lambda$, and $F_{\rm con}(\lambda)$ is the flux of pseudo-continuum and is derived from linear interpolation of fluxes on both sides of selected line bandpass \citep{1994Worthey}. 
Due to high effective temperatures for early-type stars, 
not many line features are shown in the LAMOST-MRS spectra.
Consequently, we measured the equivalent widths (EW) of the H$\alpha$ $\lambda$6564\ \AA, He I $\lambda$6679 \AA, and Mgb series \citep{2015LiuChao,2017LiuChao}.
Table~\ref{Tab:1} shows the information on index bandpass and the two continua bands for these lines \citep{1998Cohen,2015LiuChao}. Fig.~\ref{fig:All candidate} shows the distributions for our initial sample as the black points in ($EW_{\rm Mgb}$, $EW_{\rm H\alpha}$) and ($EW_{\rm He I}$, $EW_{\rm H\alpha}$) panels, respectively.

\begin{table*}
	\centering
	\caption{The definition of equivalent width}\label{Tab:1}
	\begin{tabular}{  l  c   c }
	\toprule
	\hline \\
	Lines & Index bandpass[\AA]           &Pseudo-continua[\AA] \\
	\hline \\	
	H$\alpha^a$ &   6548.00-6578.00 & 6420.00-6455.00 6600.00-6640.00\\         
	\hline \\	
	He I    &   6672.00-6689.00 &   6653.00-6672.00 6690.00-6712.00\\
	\hline \\
	$\rm Mgb^b$ &   5160.13-5192.63   &  5142.63-5161.38 5191.38-5206.38\\
	\hline \\
$\prescript{a}{}\,$\cite{1998Cohen},\\
$\prescript{b}{}\,$\cite{2015LiuChao}\\
	\end{tabular}
\end{table*}

\begin{figure*}
        \centering
	    \includegraphics[scale=0.5]{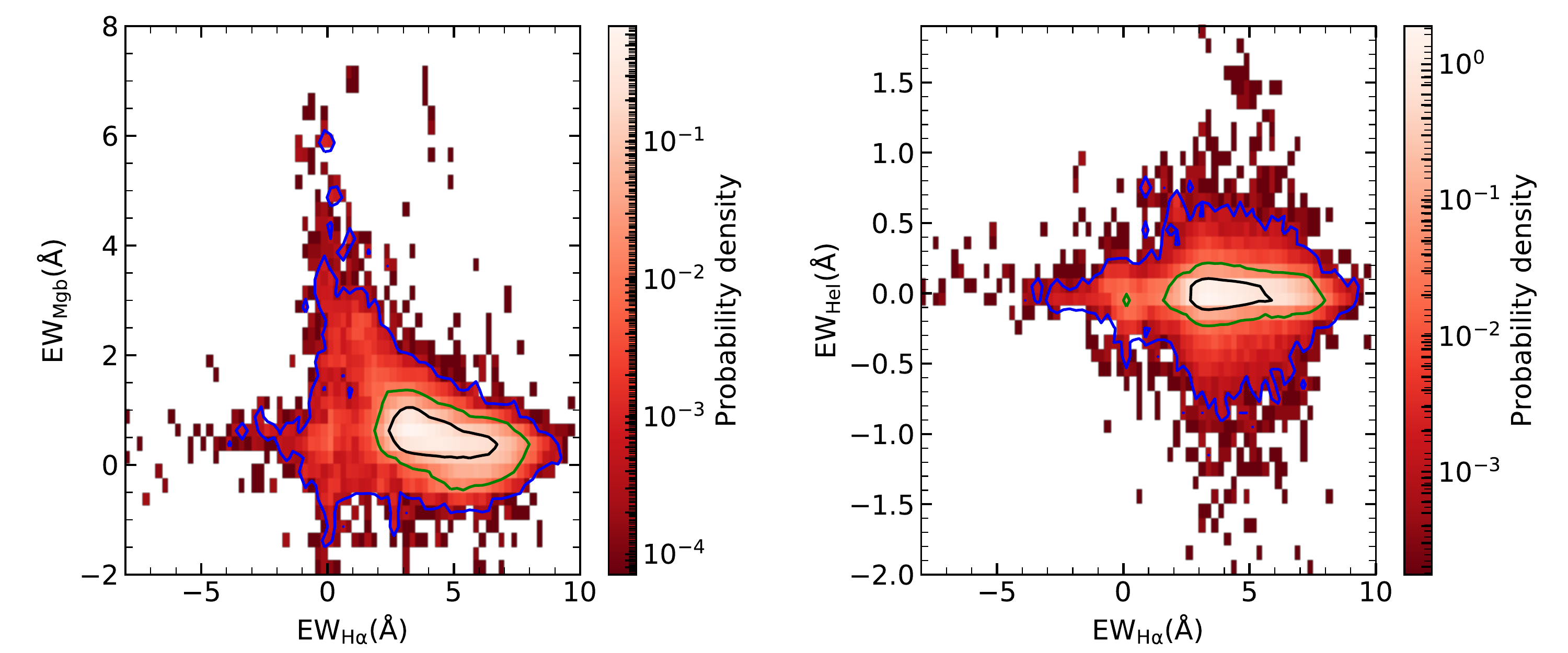}		
		\caption{The distribution of late A- and F-type (LAF) type stars in H$\alpha$ vs.\,Mgb panel (left) and H$\alpha$ vs.\,He I panel (right) in which color-bar represents the probability density of the distribution. The black and green (blue) contour lines represent the contours of stars in 1$\sigma$ and 2$\sigma$ (3$\sigma$).}\label{fig:F sigma}
\end{figure*}

As the first step, we removed late A- and F-type (LAF) stars from the sample. 
To do this, we chose LAF stars in LAMOST-MRS i.e. with $6,500 K\le T_\mathrm{eff}\le 8,000K$ \citep{1981Habets} and calculated the EWs of ${\rm H}\alpha$, HeI and Mgb, for these stars. 
Fig.~\ref{fig:F sigma} shows the probability density distributions of the EWs in ($EW_{\rm Mgb}$, $EW_{\rm H\alpha}$) plane (left panel) and in ($EW_{\rm He I}$, $EW_{\rm H\alpha}$) plane (right panel), respectively. 
The color bar on the right side indicates the probability density. 
The distribution of the LAF stars within the 1$\sigma$ (68.27\%, black),  2$\sigma$ (95.45\%, green), and 3$\sigma$ (99.73\%, blue) regions are shown with black, green and blue lines. 

In principle, we can remove most of LAF stars from the sample if we discard the objects located in the region enclosed by the 2$\sigma$ line.
However, if we do like this, we will lose a majority number of early A-type stars in our sample,
since the spectral features of F-type stars are similar to that of A-type stars, 
resulting in EW density distributions of F-type stars indistinguishable from that of A-type stars. 
This can be clearly seen in Fig.1 where we overlapped LAF stars within the region of 1$\sigma$ and 2$\sigma$.   
So, we only eliminated LAF stars within the region enclosed by 1$\sigma$ line in this step\footnote{The region enclosed by the 1$\sigma$ line corresponds to 6,978 stars shown as the blue dots of Fig.~1, while 2$\sigma$ encloses 135,732 stars as shown in the green dots of Fig.~1.}. 
We will double-check our final sample by visually checking the individual spectrum to exclude any spurious late-type stars spectrum after using all the criteria.  

\begin{figure*}
        \centering
	    \includegraphics[scale=0.5]{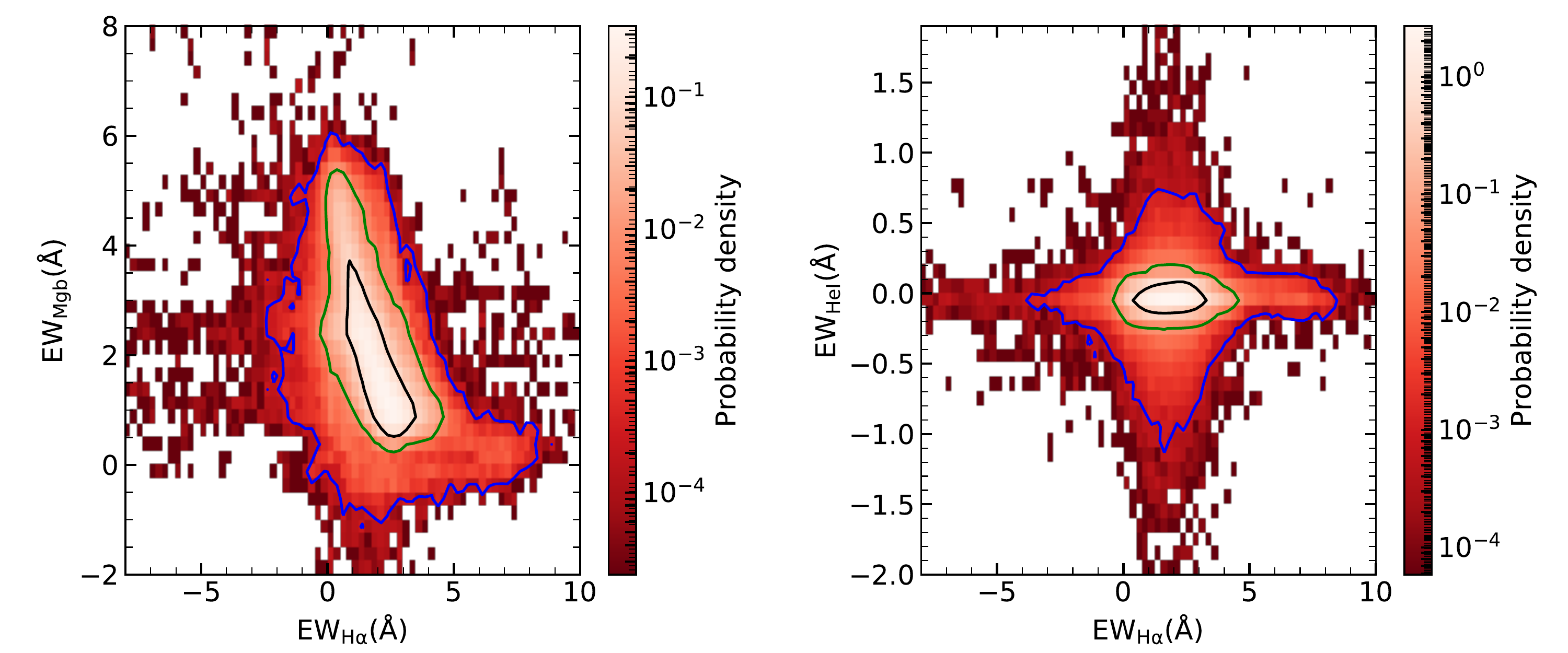}	
		\caption{The distribution of G-type stars in H$\alpha$ vs.\,Mgb panel (left) and H$\alpha$ vs.\,He I panel (right) in which color-bar represents the probability density of the distribution. The black and green (blue) contour lines represent the contours of stars in 1$\sigma$ and 2$\sigma$ (3$\sigma$).}\label{fig:G sigma}
\end{figure*}

We applied a similar approach to exclude G-type stars with $T_\mathrm{eff}$ in the range of 5200 $-$ 6500\ K \citep{1981Habets}. The probability density distributions for the G-type stars are shown in Fig.~\ref{fig:G sigma}. Stars within 2$\sigma$, corresponding to the 95.45\% confidence level regions of the distribution, were excluded from the sample.

Because LAMOST-MRS pipeline does not assign the $T_\mathrm{eff}$ to stars cooler than 3100\ K, we are not able to apply the above method to remove M- and K-type stars. We cross-matched the stars in our sample using their coordinate designation with 2$MASS$ database \citep{2003Cutri2mass} (within 3 arc seconds), and obtained the reddening-free quantity $Q$ ($Q = (J-H)-1.70(H-K_{S})$) for the common stars. \cite{2007Negueruela} suggests that for M- and K-type stars, the $Q$ values are expected to be within a range of 0.4 to 0.5. Therefore, we selected stars satisfying this condition (shown as the red dots in Fig.~\ref{fig:All candidate}) to remove them from our sample. 

Since He I line profile appears prominent in spectra of early-type stars, the line profile gradually decreases in strength towards cooler stars and is mostly absent in the spectra of A-type stars. 
We thus selected the stars with measured $EW_{\rm He I} \le 0.1$ to remove any spurious late-type stars further. We use the information of $T_\mathrm{eff}$ given by {\tt{Gaia DR2}}\footnote{\cite{2018GaiaTpipline} use an empirically-trained data-driven method to give the $T_\mathrm{eff}$ of $Gaia$ DR2. Due to the limit of training data, $T_{\rm err}$  is in the range of 3,000K - 10,000K, and the typical accuracy is 324K.} to remove any missing late-type stars in the sample. We cross-matched the data set within 3 arc seconds with $Gaia$ DR2 to find common stars and then remain stars with estimated $T_\mathrm{eff}\ge7300$K (F$_{0}$) to be our final sample of early-type stars \citep{1981Habets}\footnote{Although there are some arguments on the effective temperature given by Gaia, especially for OB stars, we only use them to remove late type stars i.e., with $T_{\rm eff}\le 7,300 K$. The real temperature is not used here and has little effect on our results.}.

Lastly, through visual inspection of the individual spectrum in the sample, we found a residual contamination rate of late-type stars about 1.6\% in our sample. We then removed these stars, as well as spectra displaying emission profiles from the sample. In our final dataset, we collected a sample of 9,382 stars with the signal-to-noise ratio (SNR) \textgreater \ 40. All the targets are with a simple magnitude cut-off G $\sim$13 \citep{2020LiuChao}.

\subsection{The sample grouping}\label{Sec:temp}
\begin{figure*}
        \centering
	    \includegraphics[scale=0.3]{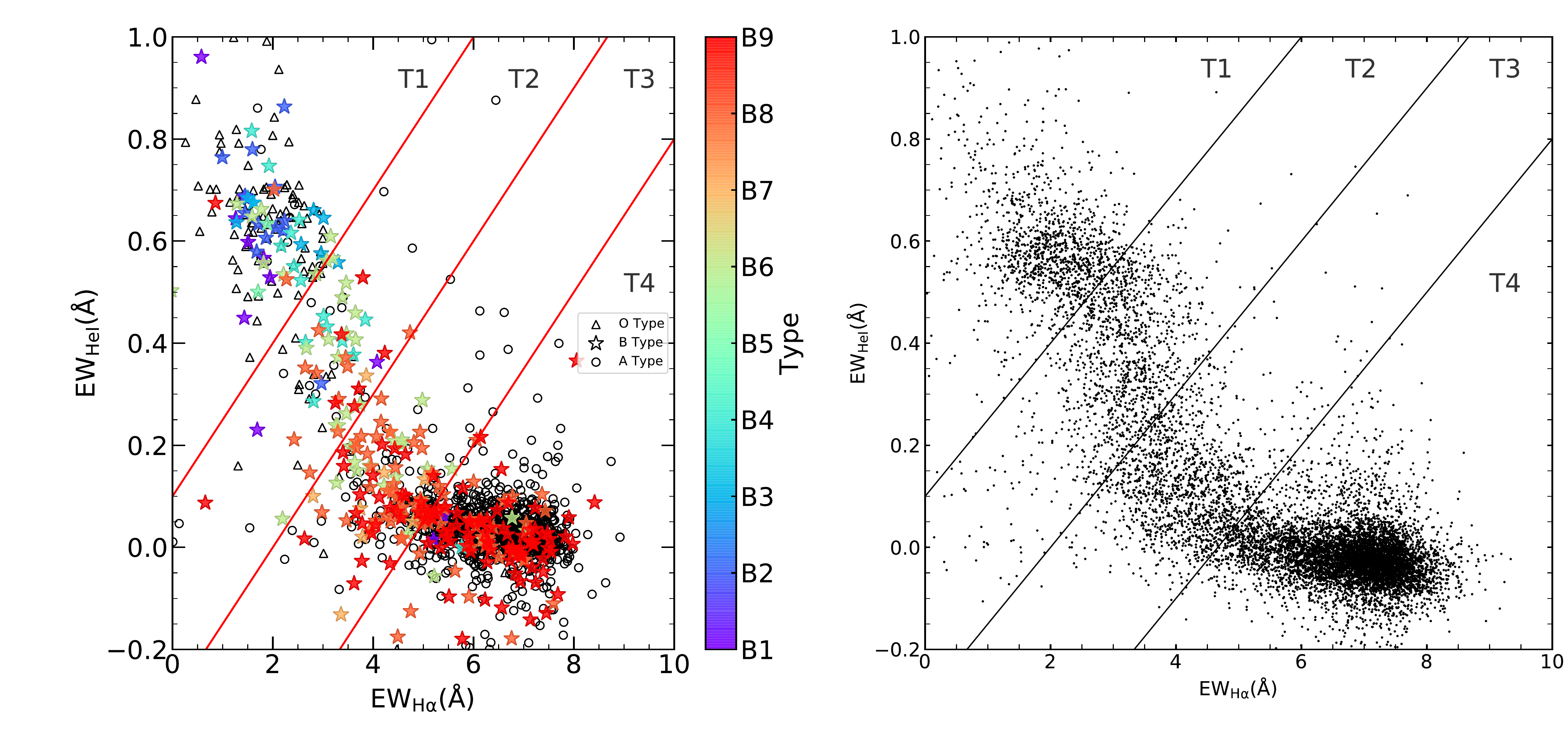}	
		\caption{Left panel: Distribution of the OBA type stars referenced by previous literatures in He I (6679\AA) vs.\,H$\alpha$ (6565\AA) panel. The black triangles and circles indicate the O- and A-type stars, respectively. The colorful stars indicate the different sub-types of B stars. The red lines indicate the boundaries between the groups. Right panel: The distribution of our final OBA candidates in the four groups of LAMOST-MRS data. The grey lines are the boundaries of the groups.}\label{fig:last}
\end{figure*}

In order to investigate the relationship of the multiplicity properties of early-type stars as a function of $T_\mathrm{eff}$, we first need to group the OBA stars in our sample based upon their $T_\mathrm{eff}$. However, no information of the $T_\mathrm{eff}$ for these stars is available. Also, it is difficult to distinguish the spectra of late O-type stars from early B-type stars due to their color degeneracy \citep{2004MaizApellaniz} (or late B-type from early A-type stars \citep{2015LiuChao}). \cite{2019LiuZhicun} has demonstrated the applicability of using the spectral line indies to classify the spectral types of early-type stars in LAMOST-LRS. We thus adopted their approach to divide the stars in our sample into four catalogs in $T_\mathrm{eff}$. We collect a sample of template stars with known spectral types from literature \citep{1956Hiltner,1995Nesterov,2003Negueruela,2010Beerer,2012Comeron,2015HouwenAcite,2019LiuZhicun}. We cross-matched these stars with the LAMOST-MRS database to collect spectra of common stars, including 136 O-type stars, 308 B-type stars, and 898 A-type stars. We then measured their EWs of the H$\alpha$ $\lambda$6564\ \AA\ and He I $\lambda$6679 \AA\ line profiles. In Fig.~\ref{fig:last}, on the left panel, we plot the measured EWs in ($EW_{\rm He I}$, $EW_{\rm H\alpha}$) plane, the triangle, and circle indicate the EW measurements for the O- and A- stars, respectively, and the star symbol represents for the B-type stars. The vertical color bar on the right side represents the spectral sub-type distribution of B-type stars. Based upon the EWs, we group the stars into four groups of T1, T2, T3, and T4. As shown in the right panel of Fig.~\ref{fig:last}, the stars in group T1 comprise mostly O $-$ B4-type stars, most B5-type stars reside in T2 group, while the majority of B7-type stars are distributed in the T3 region. Cooler stars of B8 $-$ A4 are illustrated in the T4 region on the plot \citep{2021GYJ}. On the right panel, we adopted the criteria to group our observed OBA stars in the sample.

\cite{2014Sota} presented a spectral classification for a catalog of 448 Galatic O-type stars (GOSSS). We cross-matched the LAMOST-MRS catalog with GOSSS, and only four non-emission stars were found. We also cross-matched the LAMOST catalog with the published 146 OB stars from \cite{2007Kiminki30}, only 14 common sources were found. All the 18 (4+14) sources above with the spectral type from O5.5 to B1 are in the T1 group indicating that it is reasonable to adopt the method to obtain the four groups.

In Tab.~\ref{Tab:OBA Catalogs_EW}, we list the observation ID for each of the star, their coordinates, MJD date, SNR, the associated EW measurements for all three line profiles, their classified group index, star name and V magnitude (from SIMBAD). The number of stars within each group is listed in Tab.~\ref{Tab:OBA Numbers}.

\begin{table*}
	\centering
	\caption{The OBA Catalogs}\label{Tab:OBA Catalogs_EW}
	\begin{tabular}{ccccccccccc}
	\toprule
	\hline \\
	OBSID  &   RA   & DEC  &  Date  &  $S/R$ & $EW_{Mgb}$ & $EW_{HeI}$ & $EW_{H\alpha}$ & Group & Star Name & V\\
	Name   &  (deg) & (deg) & (MJD) &            &      (\AA)         &        (\AA)      &      (\AA)    &   Name & &\\
	\hline \\
	611107178 & 0.2019 & 61.0471 & 58090.5219329 & 73 &  -0.2717 & 0.5313 & 2.6843 & 1 & LS   I +60   65 & 11.51\\
	611112023 & 0.3873 & 63.5472 & 58090.5219329 & 45 &  -0.3312 & 0.5465 & 2.3036 & 1 & LS   I +63   28 & 11.62\\
	611107142 & 0.8842 & 60.9830 & 58090.5219329 & 74 &  -0.2062 & 0.5940 & 2.5774 & 1 & LS   I +60   75 & 10.80\\
	611006035 & 0.8905 & 61.9049 & 58090.4775810 & 93 &  0.0059  & 0.8916 & 4.6431 & 1 & TYC 4018-1266-1 & 10.66\\
	611107126 & 1.0108 & 60.8488 & 58090.5219329 & 88 &  -0.1561 & 0.5865 & 2.6980 & 1 & BD+60  2664     & 10.62\\
	611007126 & 1.0123 & 60.8458 & 58090.4775810 & 90 &  -0.1564 & 0.5788 & 3.0272 & 1 & BD+60  2664B    & 10.65\\
	609206246 & 1.0765 & 55.8749 & 58088.4615394 & 92 &  -0.1816 & 0.5952 & 2.5991 & 1 & LS   I +55      & 10.76\\
	611113112 & 1.0887 & 63.0636 & 58090.5219329 & 60 &  0.2635  & 0.4098 & 2.0286 & 1 & LS   I +62      & 11.56\\
	\hline \\
	\end{tabular}\\
	Note: The full data are available in the machine-readable table online.
\end{table*}

\begin{table}
	\centering
	\caption{Numbers of OBA stars and the time-domain observation stars}\label{Tab:OBA Numbers}
	\begin{tabular}{ccc}
	\toprule
	\hline \\
	Class & No. of stars & No. of time-domain stars\\
	\hline \\	
	T1&1138&499\\
    \hline \\	
	T2&1092&366\\
	\hline \\	
	T3&2649&816\\
	\hline \\	
	T4&4503&1230\\
	\hline \\
	Total  stars&9382&2911\\	
	\toprule
	\end{tabular}
\end{table}

\section{Radial Velocity measurements and analysis}\label{sec:RV}
\subsection{RV measurements}\label{sec:Xiong RV}

\begin{table*}
	\centering
	\caption{The RV Catalogs}\label{Tab:OBA Catalogs}
	\begin{tabular}{lcccccccc}
	\toprule
	\hline \\
	OBSID  &   RA   & DEC  &  Date  &  $S/R$ &  $RV_{\rm rel}$      &     $\sigma$         &   Group \\
	Name   &  (deg) & (deg) & (MJD) &            &    (km s$^{-1}$)       &   (km s$^{-1}$)     &    Name \\
	\hline \\
	596906246 & 1.0765&55.8749 & 58058.6294676 & 73 & 3.35  &   1.08 &1\\
	599606246 & 1.0765&55.8749 & 58065.5797569 & 58 & 0.63  &   1.17 &1\\
	609206246 & 1.0765&55.8749 & 58088.4802083 & 85 & -1.15 &   0.94 &1\\
	596906156 & 2.2003&55.6665 & 58058.6294676 & 65 & 2.60  &   1.28 &1\\
	599606156 & 2.2003&55.6665 & 58065.5797569 & 74 & 6.77  &   1.19 &1\\
	609206156 & 2.2003&55.6665 & 58088.4802083 & 33 & 12.72 &   2.29 &1\\
	701403193 & 5.4396&58.3395 & 58466.4750463 & 34 & -0.08 &   1.10 &1\\
	702103193 & 5.4396&58.3395 & 58468.5096412 & 186& 0.01  &   0.27 &1\\
	\hline \\
	\end{tabular}\\
	Note: The full data are available in the machine-readable table online.
\end{table*}

\begin{figure}
        \centering
	    \includegraphics[scale=0.3]{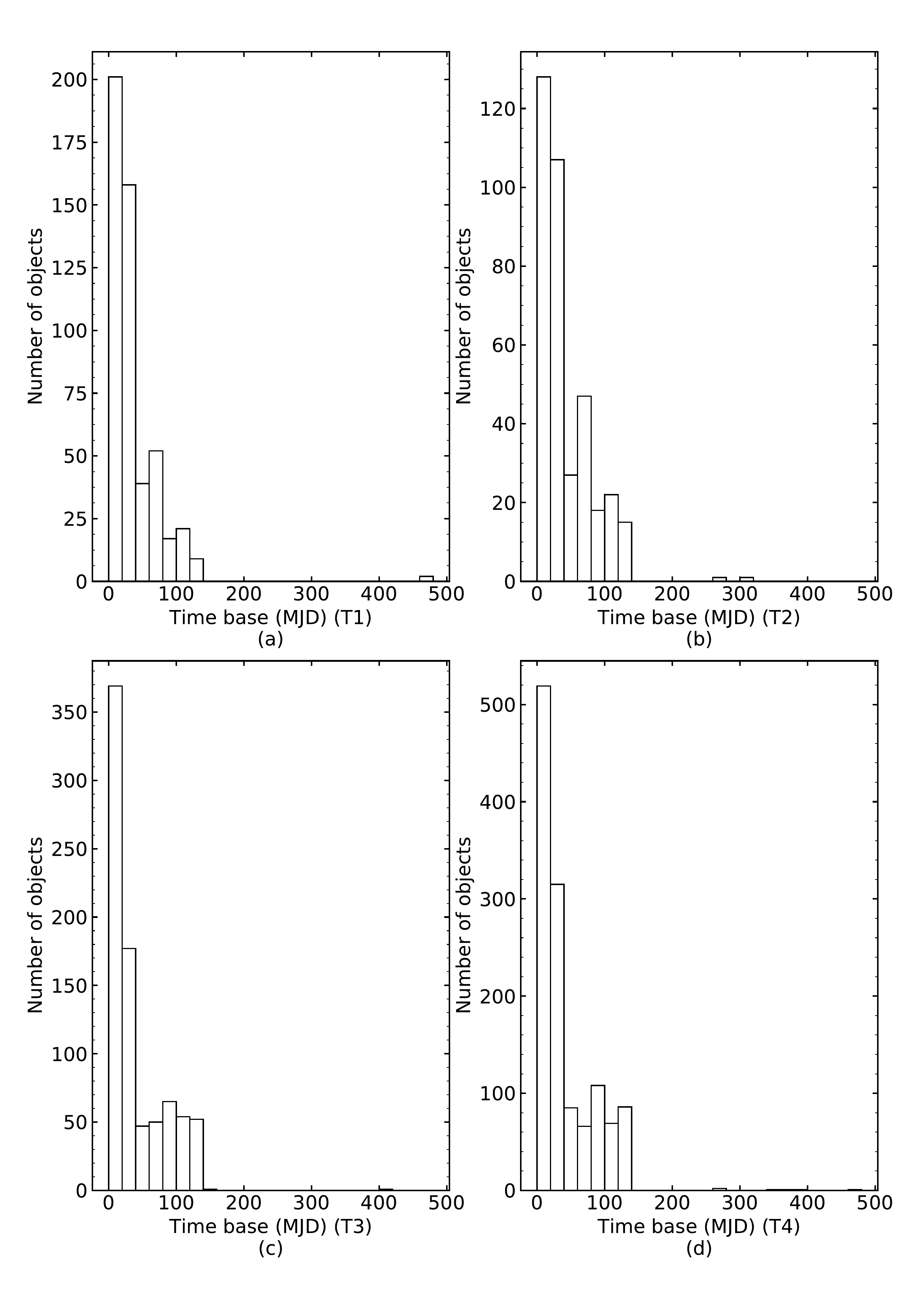}	
		\caption{The number distribution of stars as a function of observational baseline. Panels (a), (b), (c) and (d) represent the stars in groups of T1, T2, T3, and T4, respectively.}\label{fig:typical time}
\end{figure}

\begin{figure}
        \centering
	    \includegraphics[scale=0.3]{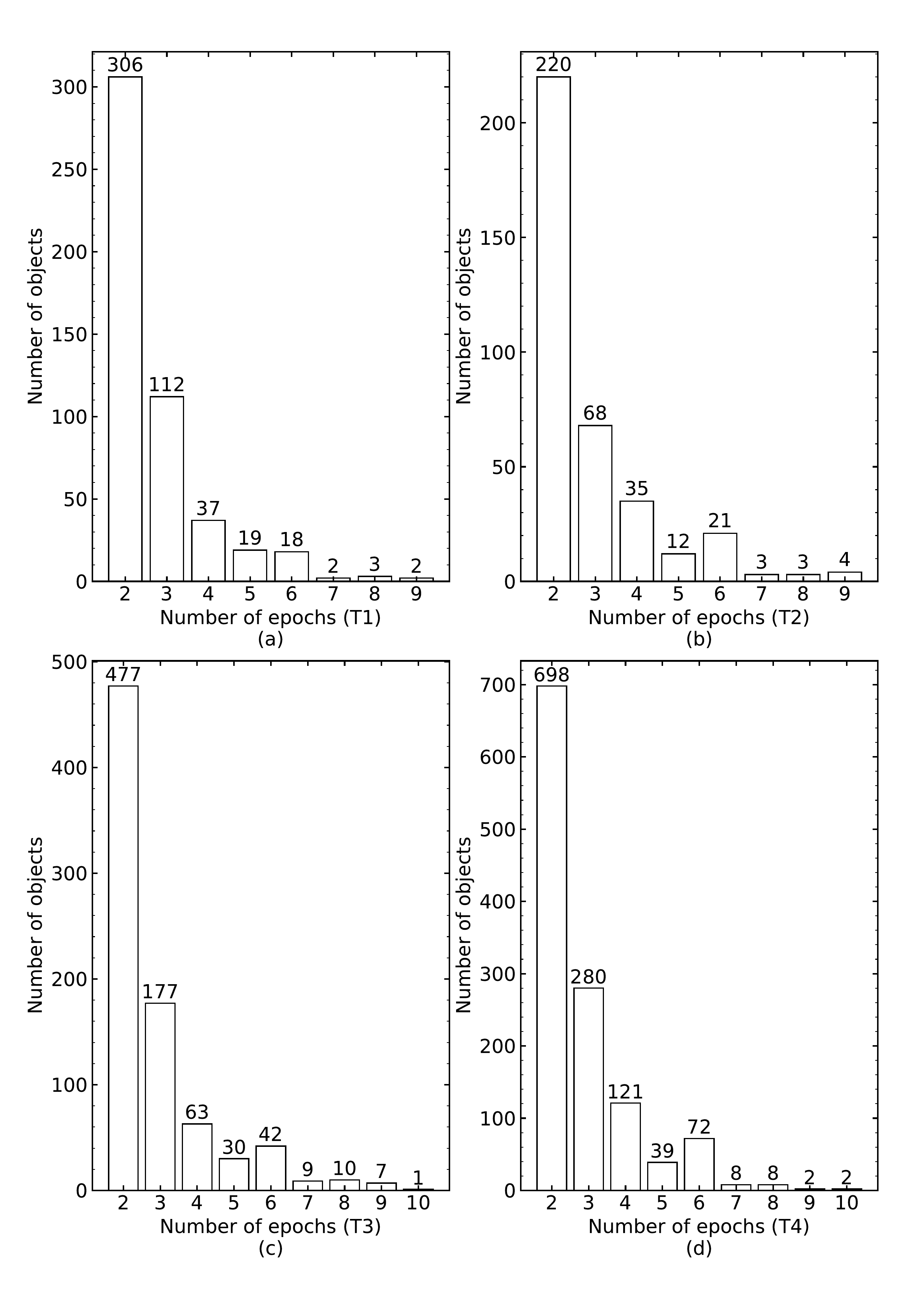}
		\caption{Number distribution for observation of SNR \textgreater 40. Panels (a), (b), (c) and (d) represent the stars in groups of T1, T2, T3, and T4, respectively. It indicates that 59\% of stars have two observations, and 10\% have five or more observations.}\label{fig:obs_times}
\end{figure}

The radial velocity measurement is a vital ingredient of this study since detecting significant RV variations relies on RV differences. We adopted the maximum likelihood method described in \citep{2021XiongJianping} to measure the relative RVs. We briefly described the procedures as follows. For each star, a set of multiple exposures were obtained from the LAMOST-MRS, and we selected the exposure with the highest SNR as the template spectrum. The likelihood distribution for the $i$th star observed at time $t$ of the relative radial velocity ($RV_{\rm rel}$) is

\begin{equation}\label{eq:likelihood}
\centerline{$p(RV_{\rm rel})$=$\prod_{\lambda}\frac{\exp[-\frac{(f_{t,i}(v,\lambda)-f_{\rm tem,\it i}(\lambda))^2}{2(\sigma_{t,i}^2+\sigma_{\rm tem,\it i}^2)}]}
{\sqrt{2\pi(\sigma_{t,i}^2+\sigma_{\rm tem,\it i}^2)}}$}
\end{equation}
where the $f_{\rm tem,\it i}(\lambda)$ and $f_{t,i}(v,\lambda)$ represent the observed flux of template spectrum and individual exposure, respectively, while the $\sigma_{\rm tem,\it i}$ ($\sigma_{t,i}$) is the observed flux error of template spectrum (individual exposure).
A Gaussian fit was applied to each of the likelihood distribution functions, and we estimated the peak value of the fit as our $RV_{\rm rel}$ measurement, and the uncertainty ($\sigma_{\rm unc,\it i}$) of the RV measurements was obtained from the standard deviation of the Gaussian fit. Such an approach yields a systemic error of 0.25 $\rm km\,s^{-1}$ ($\sigma_{\rm sys}$). 

The systematic errors exist among the RVs obtained from spectra collected by different spectrographs and exposures of LAMOST MRS surveys \citep{2019liunianRV,2021ZhangboRV}. Therefore, we matched our catalog, including 2911 early-type stars, with the one from \cite{2021ZhangboRV} and found that the median values of the RVs before correction and after correction are $\sim$0.6km/s. In the paper, we used a strict criteria to select the binary stars (see Equation.~\ref{cer:SB1}). This means that the influence of uncorrected RV on the derived binary fraction is little and can be ignored. We also compared the observed binary fraction based on the RVs from \cite{2021ZhangboRV} and \cite{2021XiongJianping} and found that the difference between the observed binary fraction deriving from the two way is less than 1.5\%. Therefore, we kept using the $RV_{\rm rel}$ from \cite{2021XiongJianping} in the following analysis.

In our final sample of 9,382 candidates, only 2,911 stars have been observed more than twice (see column 3 in Tab.~\ref{Tab:OBA Numbers}).
The following analysis of this work is based on the sample of 2,911 stars. 
The measured $RV_{\rm rel}$ and the uncertainties ($\sigma_{\rm unc,\it i}$) for each star are tabulated in Tab.~\ref{Tab:OBA Catalogs}. The number distribution of the stars as a function of observing baseline is shown in Fig.~\ref{fig:typical time}. In Fig.~\ref{fig:obs_times}, we displayed the number of observations for stars in each temperature group.

\subsection{Criterion for binarity}\label{sec:Criteria}

\begin{figure*}
        \centering
	    \includegraphics[scale=0.5]{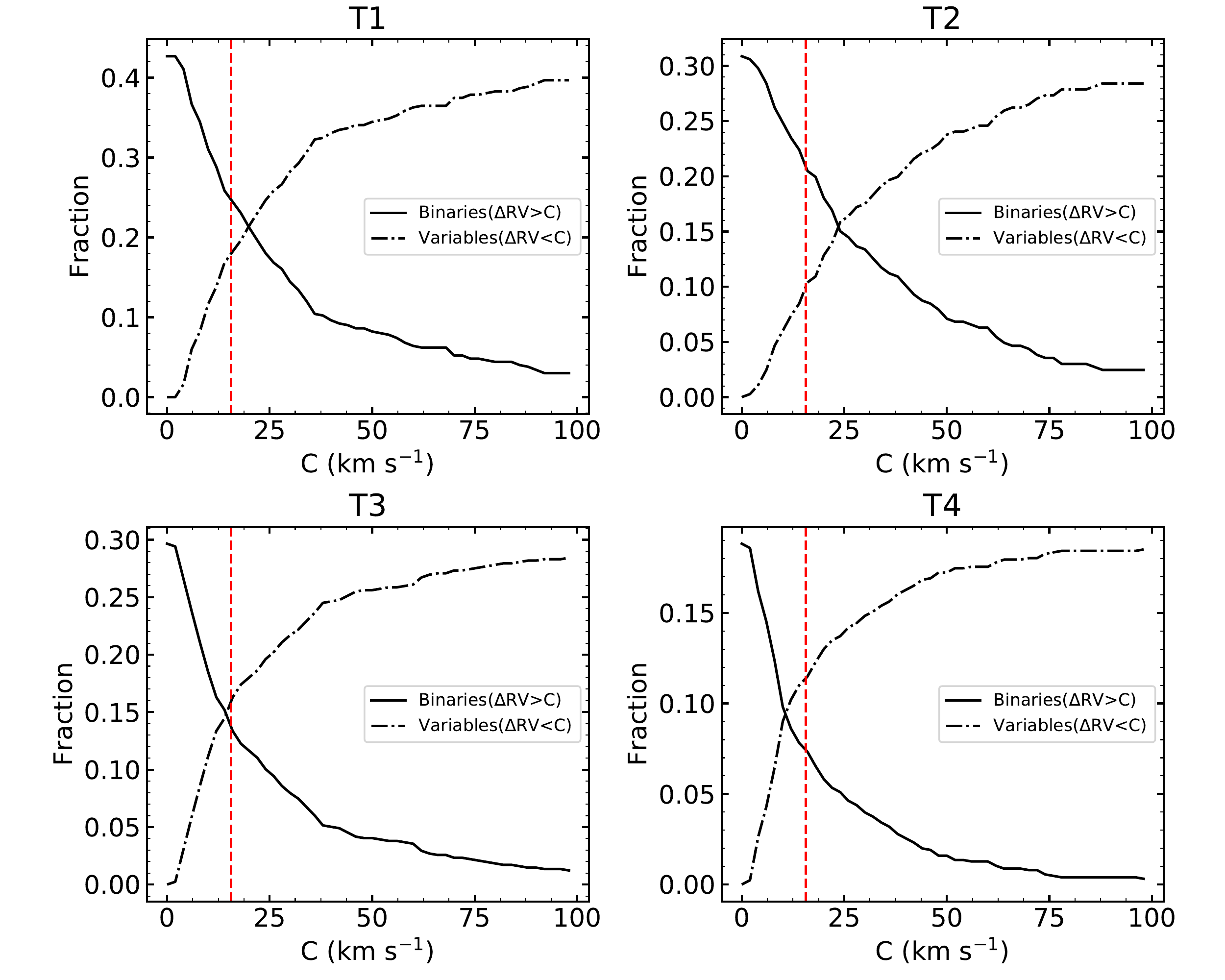}
		\caption{The varied fraction of systems below and above the critical C. The solid line indicates the varied fraction of systems with the different thresholds of C. The dashed-dotted lines indicate the low-amplitude RV variable fractions of systems with the different thresholds of C. The vertical red dashed lines represent the final adopted threshold of C = 15.57 $\rm km\,s^{-1}$.}\label{fig:kink}
\end{figure*}

We adopted the method from \cite{2013Sana} (see their Equation 4) to identify binary candidate systems in our sample \citep{2012Sana1,2015Dunstall,2021MahyC,2021Banyard}. 
We considered a star as a binary if $RV_{\rm rel}$ satisfies the criterion 
\begin{equation}
\centerline{ $\frac{|v_{i} - v_{j}|}{\sqrt{\delta_{i}^2 \ + \delta_{j}^2}}\  > \ 4$ and ${|\\ v_{i} - v_{j} \\ |}\  >  \ C$,}\label{cer:SB1}
\end{equation}
where $v_{i(j)}$ and $\delta_{i(j)}$ are the $RV_{\rm rel}$ and the associated uncertainty measured for the spectrum at epoch i~(j). 
Here we divided the uncertainty into two parts:
\begin{equation}
\centerline{ $\delta_{i} \  =  \ {\sqrt{\ \sigma_{\rm unc,\it i}^2 \ + \ \sigma_{\rm sys}^2}}$,}\label{eq:RV_sigma}
\end{equation}
where $\sigma_{\rm unc,\it i}$ is the uncertainty of the $i$th star observed at time t while $\sigma_{\rm sys}$ is the typical systematic error (see Sec.~\ref{sec:Xiong RV}).

The threshold C is adopted to eliminate the stars with significant RV variation caused by pulsation of the photosphere or atmospheric activity. \cite{2013Sana} and \cite{2015Dunstall} show that this value is assigned from a kink on the RV distribution plot (see Fig.~3 in \cite{2013Sana}). We followed a similar approach of searching for possible kink from the RV distribution to constrain the threshold C value. However, no such pattern is shown in our RV plot shown in Fig.\ref{fig:kink}.
In order to eliminate the pulsational variable stars in our sample, we adopted the typical period and RV amplitude of $\delta$ Scuti type stars\footnote{
We here care about pulsating variable stars with OBA types, that is, 
$\beta$ Cephei, Slowly Pulsating B (SPB), and $\delta$ Scuti. 
We cross-matched the catalog of \cite{2017KochanekASAS}, \cite{2019Gaiavariables}, and \cite{2020chenxiaodianZTF} with our catalog, but only found five sources in common (we have eliminated them). The contamination by $\delta$ Scuti stars in our sample is much more likely than that by $\beta$ Cephei and SPB \citep{2006WatsonV, 2020chenxiaodianZTF}. 
So we adopted the threshold value derived from $\delta$ Scuti in this paper.
This value is larger than that for $\beta$ Cephei \citep{2005StankovBCEP}
and likely leads to a lower observed binary fraction. However,
we aim to remove contamination of pulsating variable stars as much as possible to ensure the purity of the sample. It has little effect on the intrinsic binary fraction after the MCMC correction.
} to perform a Monte-Carlo Simulation to constrain the threshold C value, resulting in a value of 15.57\,$\rm km\,s^{-1}$ \citep{1979Breger,2000Breger,1994Fernie}.

\subsection{Observed binary fraction}\label{sec:fb_obs}
\begin{figure}
        \centering
	    \includegraphics[scale=0.5]{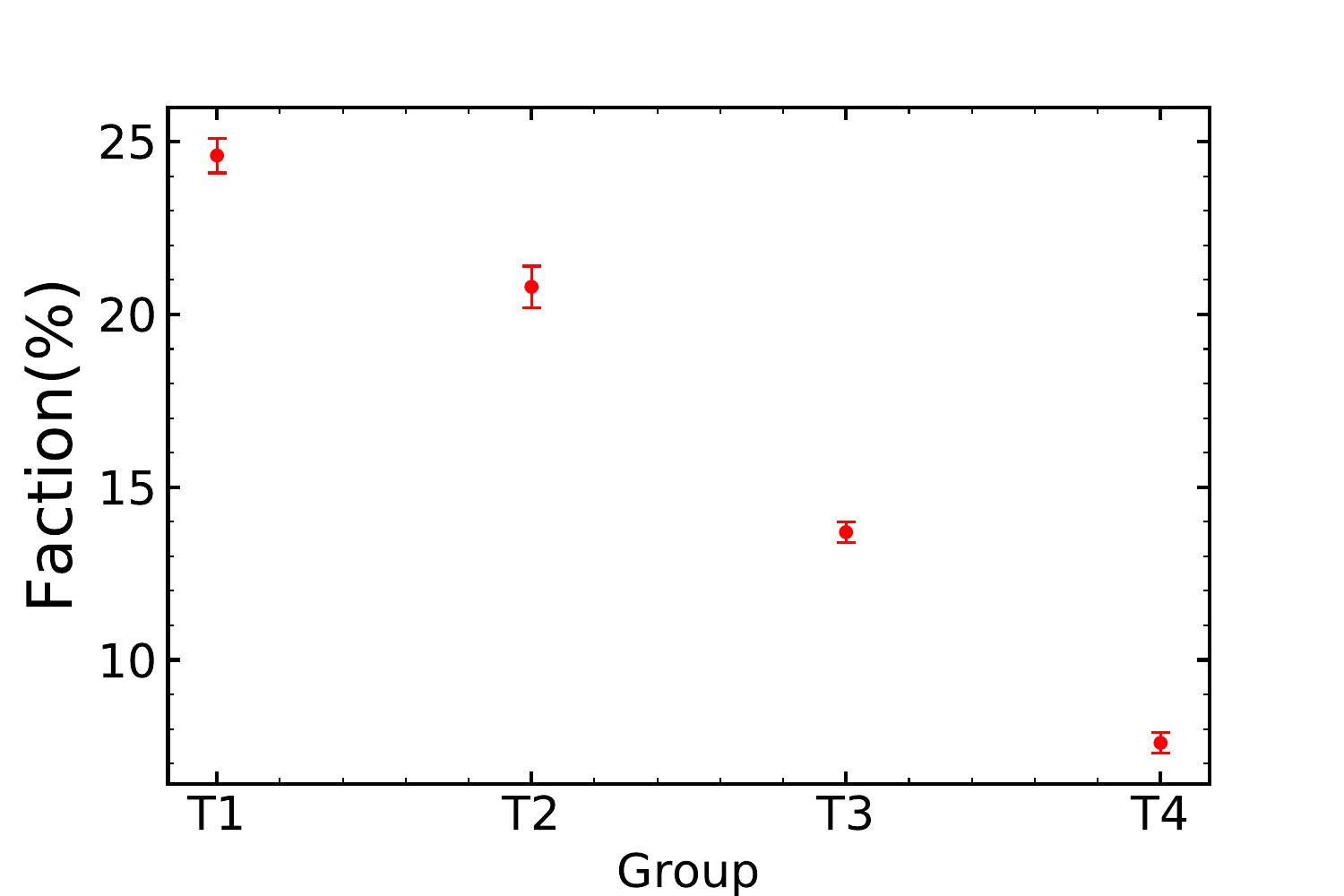}
		\caption{The observed binary factions for the four groups.}\label{fig:obs_result}
\end{figure}

According to the criterion (\ref{cer:SB1}), the observed binary factions ($f_{\rm b}^{\rm o}$) of these four groups are $24.6\%\pm0.5\%$, $20.8\%\pm0.6\%$, $13.7\%\pm0.3\%$, and $7.6\%\pm0.3\%$, respectively. The $f_{\rm b}^{\rm o}$ for the stars in each group is plotted in red dots and showed in Fig.~\ref{fig:obs_result}. The error bars were estimated through bootstrap analysis from \cite{2010Raghavan}. 
The result shows that $f_{\rm b}^{\rm o}$  decreases as the spectral type moving from O-type to A-type stars.

In fact, this way may underestimate the fraction of binary stars, since the binaries with relatively small RV variations in the measurements based on the LAMOST MRS spectra are missed. In addition, if the phase differences of the spectroscopic observations of LAMOST are small, even the binary stars with large RV amplitudes might be missed. Besides, different threshold C-value may affect the result of $f_{\rm b}^{\rm o}$, but all the bias above can be corrected when we investigate the intrinsic binary fraction through Monte-Carlo simulations in the next section. 

\section{Intrinsic multiplicity properties}\label{sec:MCMC all}
As stated above, the observed $f_{\rm b}^{\rm o}$ depends on the threshold C-value.
Besides, based upon the observational baseline and cadence nature for stars in our sample, 
we may miss the detection of long period binary systems. 
In this section, we perform a series of Monte-Carlo simulations to correct such biases.

\subsection{Monte-Carlo method}\label{sec:MCMC}
Following the approach as noted by \cite{2013Sana}, we need to adopt the Monte-Carlo simulation to construct synthetic cumulative distributions (CDF) of RV variance ($\Delta$RV, is the maximum RV variance among individual exposure) and $\Delta$MJD (minimum time scale between the exposures). We assume the probability density distribution of orbital period~($P$) and mass ratio ($q$) of binary system satisfying the power-low $f(P) \propto P^\pi$ and $f(q) \propto q^\kappa$, respectively. We modified the distribution of orbital period in log-scalar of (\rm log $P$) $\propto$ (log\ $P)^\pi$ to linear form to be more sensitive to short-period binaries. In Tab.~\ref{Tab:sim parameter}, we list the variable ranges for the parameters, and we set the step size to be 0.1 for $\pi$ and $\kappa$, and 0.04 for $f_{\rm b}$. We adopted the initial mass function \citep{1955Salpeter} to describe the mass distribution of stars in our grouped catalogs of T1 $\sim$ T4.

Generally, six parameters are used to describe two-body Kinetic systems in a Keplerian orbit: inclination (i), the argument of periastron ($\omega$), true anomaly ($\nu$), eccentricity (e), semi-major axis (a), and the epoch of periastron($\tau$). The inclination satisfies a probability distribution of $sin(i)$ and is randomly drawn over an interval from 0 to $\pi$/2. The argument of periapsis satisfies a uniform distribution and is randomly drawn from it from 0 to 2$\pi$. We use $e^\eta$ to describe the distribution and set $\eta$ as -0.5, the same as that in \cite{2013Sana}. The semi-major axis depends on the orbital periods ($P$). The $\tau$ is selected at random with the unit of days. 

We followed the same method as discussed in Appendix C of \cite{2013Sana}, and used our selected parameters with assigned range to perform the simulation to obtain cumulative distributions of $\Delta$RV and $\Delta$MJD. We followed the similar approach of \cite{2013Sana} to obtain the global merit function (GMF) using the simulated CDFs and our observations. The final results are given by the projection of the GMFs, i.e., $\pi$, $\kappa$, and $f_{b}$.

\begin{table}
	\centering
	\caption{The range of parameters used in the Monte-Carlo simulation. $P$ and $q$ are the probability density distribution of the orbital period and mass ratio, respectively. $\pi$ and $\kappa$ are the indexes of the power-low for $P$ and $q$. $f_{b}$ is the binary fraction for simulation.}\label{Tab:sim parameter}
	\begin{tabular}{ c  c}
	    \toprule
	    \hline \\
	     Parameter & Variable Range\\
	    \hline \\	
	    $P$(d)   & 1 - 1000 \\
	    \hline \\
	    $\pi$ & -2.50- 2.50\\
       	    \hline \\	
	    $q$  & 0.1 - 1.0\\
	    \hline \\
	    $\kappa$ & -2.50- $2.50^*$\\
	    \hline \\	
	    $f_{b}$ & 0.20 - $1.00^*$ \\
	    \hline \\	
	\end{tabular}\\
$\prescript{*}{}\,$As for T3 and T4, the range of $\kappa$ is -3.5 - 1.50 \citep{2015Dunstall} and $f_{b}$ is 0.04 - 0.8.
\end{table}

\subsection{Validation}\label{sec:Suitability}
In order to verify the robustness of applying the methodology from \cite{2013Sana} to our dataset, we performed a self-consistency validate test, as well as testing for the suitability of adopting the GMF. 
\subsubsection{Self-consistency test}
To verify the self-consistency of our code, we collected the reported RV measurement of 360 stars from \cite{2013Sana} and obtain the optimal $\pi$, $\kappa$, and $f_{\rm b}$ values. 
The optimal values of the distribution were found to be $\pi$=-0.4$\pm0.4$, $\kappa$=-0.9$\pm0.4$, and $f_{\rm b}$=52$\pm5\%$, while \cite{2013Sana} reported their findings of $\pi$=-0.45$\pm$0.3, $\kappa$=-1$\pm$0.4, and $f_{b}$=51\%$\pm$4\%. 
Considering the uncertainties of the parameters, 
our calculated values are in agreement with that of \cite{2013Sana}.

\begin{figure*}
        \centering
	    \includegraphics[scale=0.6]{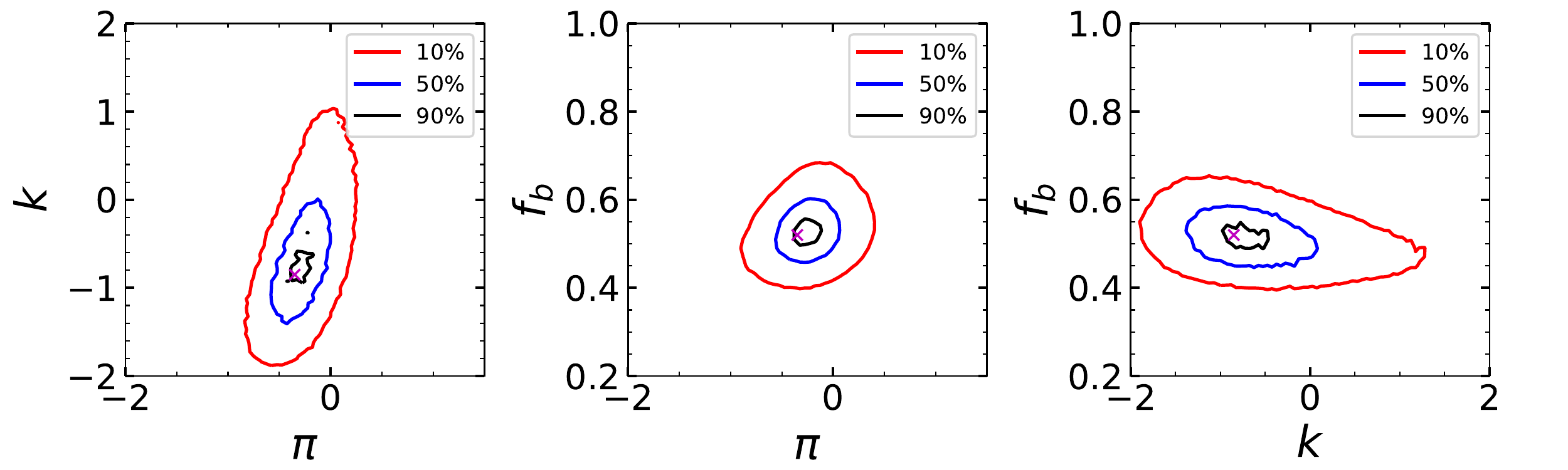}	
		\caption{Projections of GMF defined by $\pi$, $\kappa$, $f_{b}$ from the data of \cite{2013Sana}. The pink x indicates the absolute maximum. The loci of 10\%, 50\%, 90\% confidence levels of GMF are shown in the red, blue, and black contours, respectively.}\label{fig:sana_result}
\end{figure*}

\subsubsection{The applicability of the GMF}\label{sec:Suitability test}
\begin{figure*}
	\centering
	\includegraphics[scale=0.5]{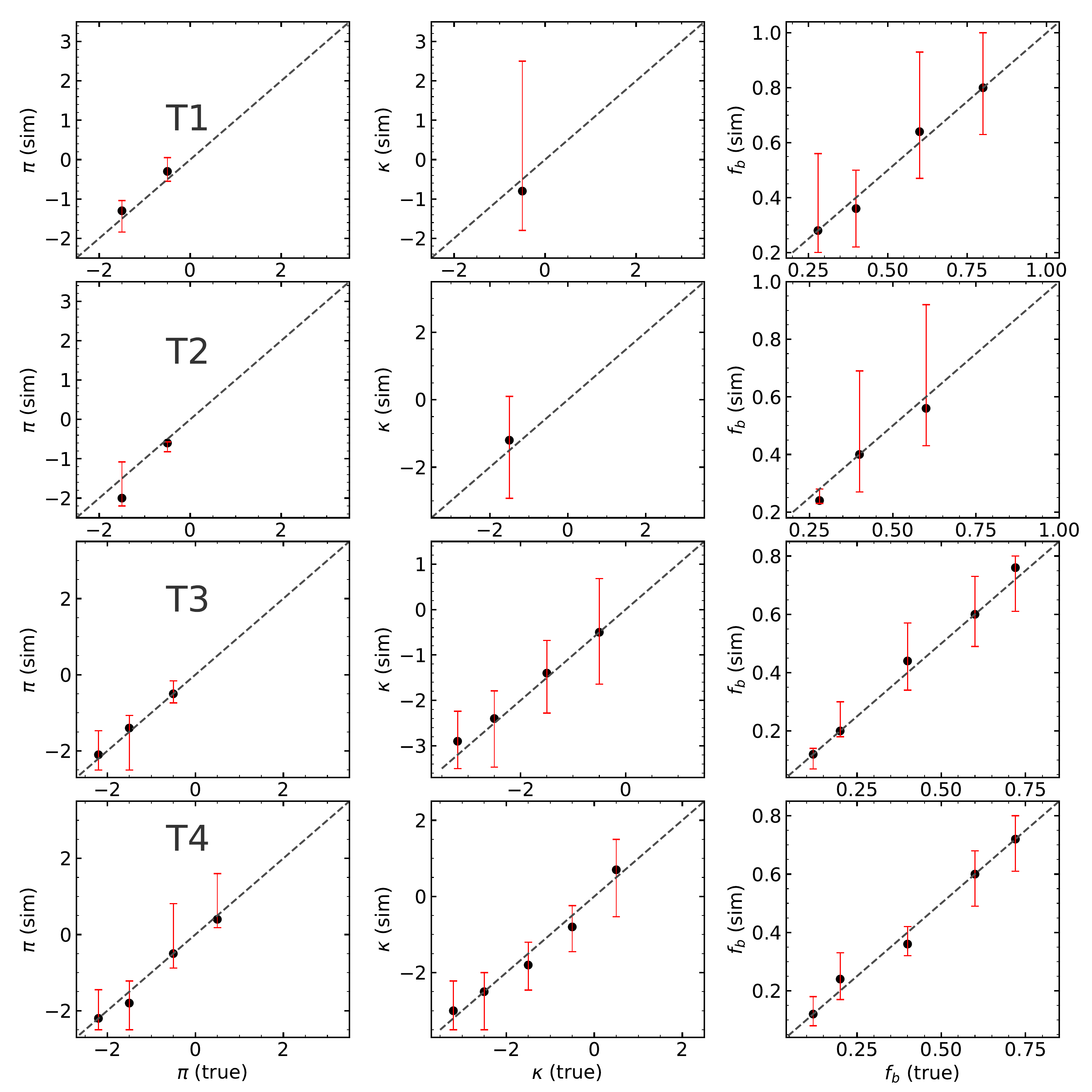}
	\caption{Comparison for testing the suitability of the global merit function. The x-axis represents the input value from Tab.~\ref{tab:choose value} while the y-axis represents the output constrained values after using the Monte-Carlo simulation, in which we did not show the points with unconstrained results. The errors are given by the upper and lower boundaries of 50\% confidence levels.}\label{fig:range_test}
\end{figure*}

\begin{figure*}
        \centering
	    \includegraphics[scale=0.6]{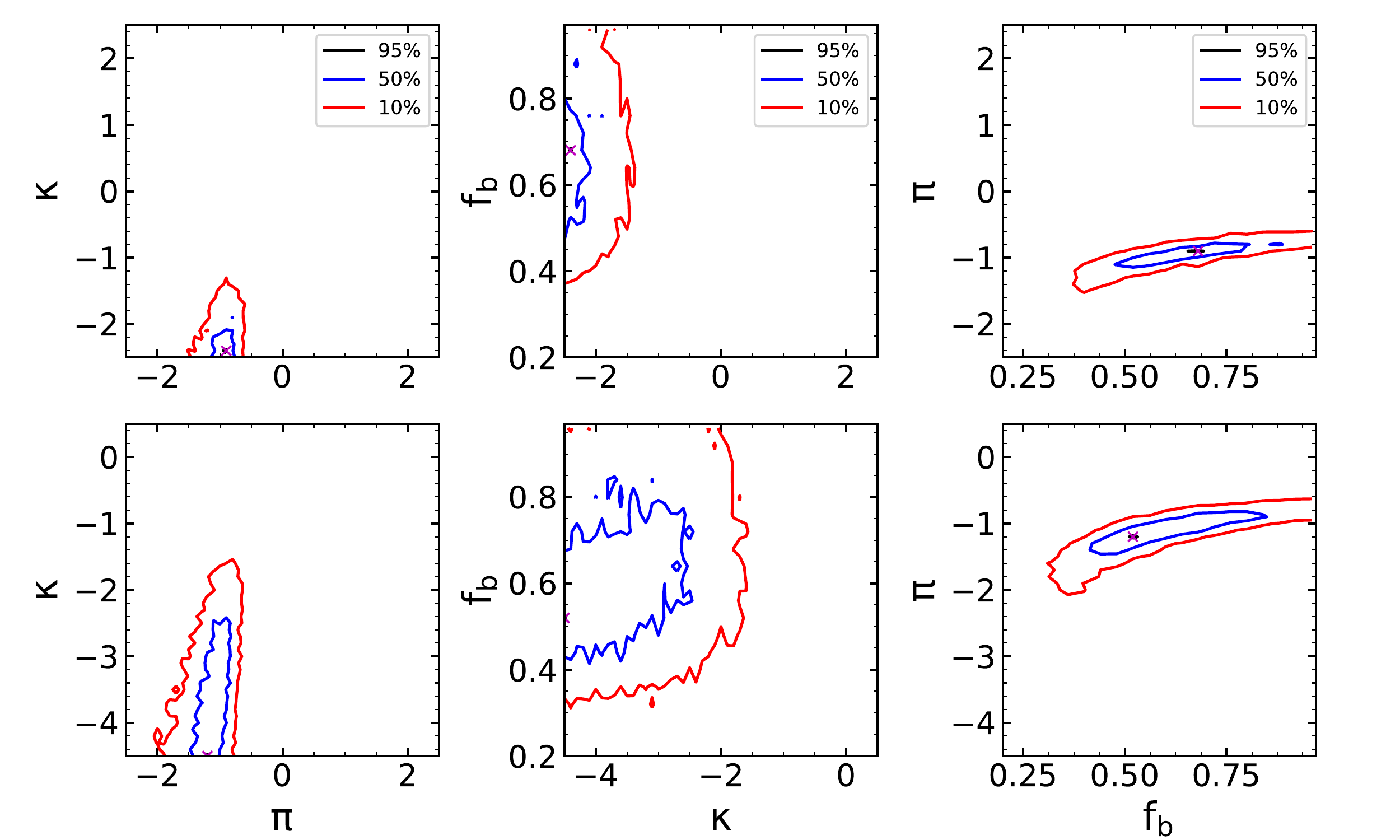}	
		\caption{The projections of GMF for stars in groups T1, but with different ranges of $\kappa$. The $\kappa$ range of [-2.5,2.5] is in the upper panel while [4.5,0.5] in the bottom panel.}\label{fig:kappa2545}
\end{figure*}

In order to test the applicability of applying the global merit function to determine the probability density distribution of sample stars from the LAMOST MRS, we first constructed 68 synthetic CDFs of $\Delta$RV and $\Delta$MJD using the input sets of $\pi$, $\kappa$, and $f_{\rm b}$. The choices of such values are listed in Tab.~\ref{tab:choose value}. Based upon the constructed input synthetic CDFs and results discussed in Sec.~\ref{sec:MCMC}, we then obtain the 68 projections of GMFs. A comparison of the output synthetic GMFs to the input parameters of $\pi$, $\kappa$, and $f_{\rm b}$ are shown in Fig.~\ref{fig:range_test}.

\begin{table*}
	\centering
	\caption{The input parameters of $\pi$, $\kappa$, and $f_{b}$. Logically, the values should be chosen with equal intervals, e.g., chosen 0.3, 0.4, 0.5, 0.6, 0.7, 0.8, 0.9 for $f_{\rm b}$, but it is time-consuming. Therefore, we selected a few values within the range but the larger interval. On the other hand, the step size for $f_{\rm b}$ is 0.04, and we selected the values of two-step size away from the boundary for the values closest to the boundary, i.e., chosen 0.28,0.92 (the same reason for the other two parameters).}\label{tab:choose value}
	\begin{tabular}{ c  c  c  c}
	    \toprule
	    \hline \\
	     Group & $\pi$ & $\kappa$ & $f_{b}$\\
	    \hline \\	
	    T1 (T2) & [-2.2,-1.5,-0.5,0.5,1.5,2.2]  & [-2.2,-1.5,-0.5,0.5,1.5,2.2] & [0.28,0.4,0.6,0.8,0.92]  \\
	            & $\kappa$=-0.5 $f_{b}$=0.6     & $\pi$=-0.5 $f_{b}$=0.6       & $\pi$=-0.5 $\kappa$=-0.5 \\
        \hline \\	
        T3 (T4) & [-2.2,-1.5,-0.5,0.5,1.5,2.2]  & [-3.2,-2.5,-1.5,-0.5,0.5,1.2] & [0.12,0.2,0.4,0.6,0.72] \\
	            & $\kappa$=-1.5 $f_{b}$=0.4     & $\pi$=-0.5 $f_{b}$=0.4       & $\pi$=-0.5 $\kappa$=-1.5 \\
	    \toprule
	\end{tabular}
\end{table*}


The top two rows show the constrained results for the groups in T1 and T2 from the synthetic CDFs distribution, while the bottom two rows are for the groups in T3 and T4. We found that the results of stars in groups T3 and T4 (12 and 14 points) are better constrained than those from T1 and T2 (7 and 6 points), in which the point represents the constrained result.
The results indicated that the method is applicable for sample stars in groups T3 and T4, but not for stars in groups T1 and T2, especially for $\kappa$. In order to explore the reasons for the unconstrained $\kappa$ value, we repeat the analysis above using the {\it observed} CDFs of $\Delta$RV and $\Delta$MJD for stars in group T1.
The results are displayed in the top panel of Fig.~\ref{fig:kappa2545}. We noted that the $\kappa$ value is still not well constrained. 
We thus further explore the possible $\kappa$ values by extending the range to -4.5 $\sim$ 0.5. The results are displayed in the bottom panel of Fig.~\ref{fig:kappa2545},
where we see that the $\kappa$ is still unconstrained. The results for T2 are similar to that of T1.  
It indicated that the initially selected range of $\kappa$ values is independent of whether it can be constrained or not.
This might be explained by the small value of sample size (N) for groups T1 and T2, which is an initial input of the Binomial distribution of the GMF. 

Considering that a large number of unconstrained $\kappa$ values for stars in the groups of T1 and T2 may affect the results of other parameters, we fixed the $\kappa$ value to guarantee the reliability of other parameters \citep{2013Sana}. According to the initial mass function, the population of late-type stars shows an increasing trend towards the small stellar mass. Therefore, the sample size of early-B type stars is large than that of the O-type stars for the stars in groups T1 and T2. We thus adopt the value of $\kappa$ from \cite{2015Dunstall} i.e. $\kappa=-2.8$ for early B-type stars and repeat the analysis above again to verify the applicability of the method. 
Based upon the validation tests, we then obtained the final results of $\pi$ and $f_{\rm b}$ from the GMFs by repeating the procedures as mentioned in Sec.~\ref{sec:MCMC}, except that we now fixed the value of $\kappa$ to be -2.8 for sample stars in groups T1 and T2.

\subsection{Results and Discussion}
\subsubsection{Results}
\begin{figure*}
        \centering
	    \includegraphics[scale=0.6]{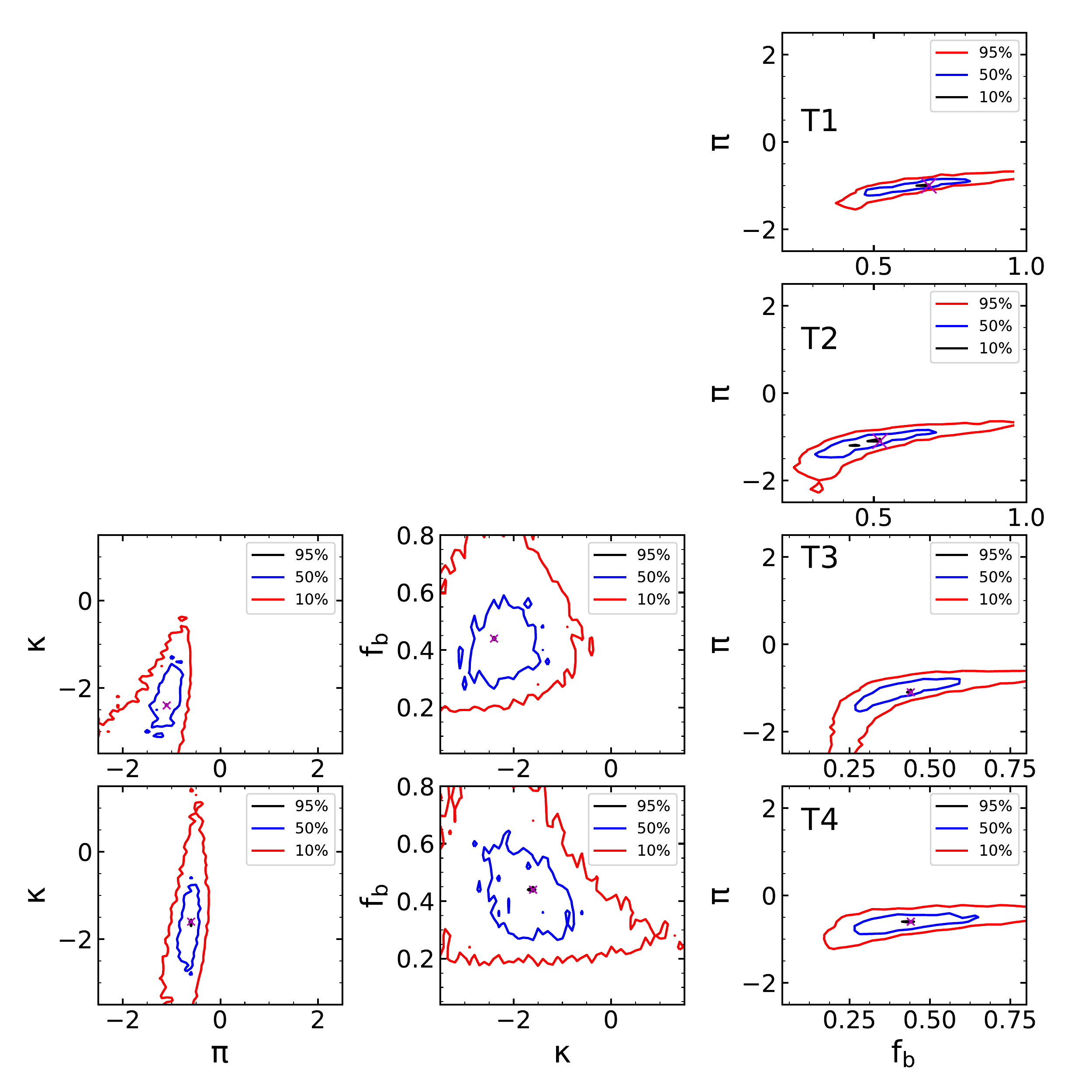}	
		\caption{The projections of GMFs for stars in groups T1 to T4. The left and middle (right) panels are the projections of $\pi$ and $\kappa$ ($f_{b}$) for stars in group T1 to T4.}\label{fig:sim_result2}
\end{figure*}

The GMFs projections of $\pi$, $\kappa$, and $f_{b}$ for stars in groups of T1 to T4 are shown in Fig.~\ref{fig:sim_result2}. 
The red, blue, and black contours indicate the loci of 10\%, 50\%, 95\% confidence levels of GMF. 
The intrinsic binary fractions of these four groups are 68$\%\pm8\%$, 52$\%\pm3\%$, 44$\%\pm6\%$, and 44$\%\pm6\%$, respectively. The $\pi$ of these groups are -1$\pm0.1$, -1.1$\pm0.05$, -1.1$\pm0.1$, and -0.6$\pm0.05$, respectively. The $\kappa$ values of T3 and T4 are -2.4$\pm0.3$ and -1.6$\pm0.3$, respectively. 

Note that, although the threshold C-value does affect the observed binary fraction $f_{\rm b}^{\rm o}$, the bias has been corrected here for the intrinsic fraction since the same C-value is used in our Monte-Carlo simulations.
\cite{2021MahyC} analyzed different threshold C-values using the Monte-Carlo method from \cite{2013Sana} to see whether different choices of C-value affect the final estimation for the intrinsic binary fraction or not. They found that the adopted threshold C does not affect the final results unless significant contamination exists by false-positive RV measurements. 

\subsubsection{Discussion}\label{Discussion}
\begin{figure}
        \centering
	    \includegraphics[scale=0.5]{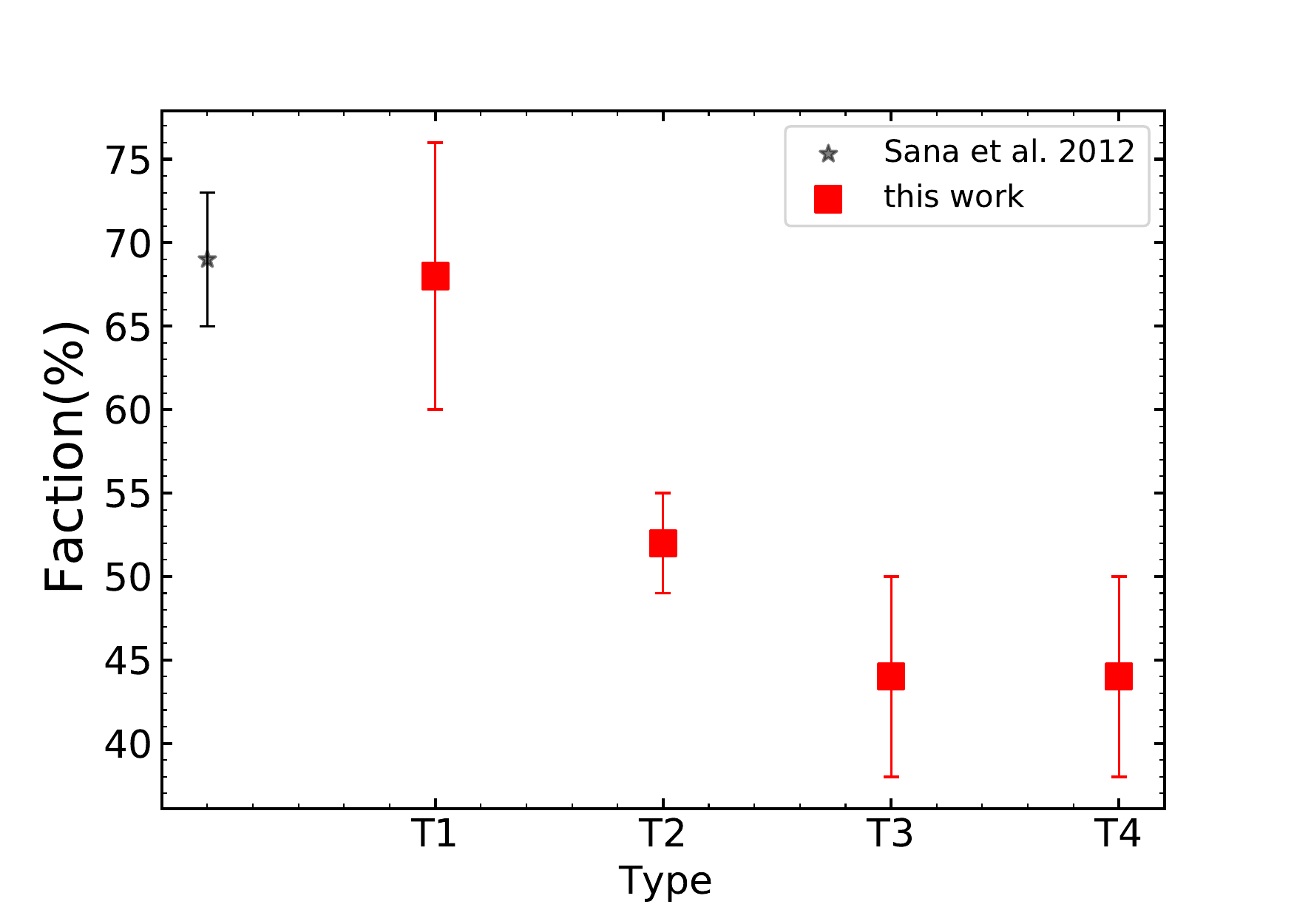}	
		\caption{Comparison of our intrinsic binary factions and previous works. The red squares represent the intrinsic binary fractions in our work for stars in groups T1 to T4, and the star mark is the result of Galactic O-type stars from \cite{2012Sana1}.}\label{fig:sim_result}
\end{figure}

Our intrinsic binary fractions (red squares) for stars in groups T1 to T4 are shown in Fig.~\ref{fig:sim_result}, and the star mark is the result of Galactic O-type stars from \cite{2012Sana1}. The final result indicated that the intrinsic binary fractions decreased toward the late-type stars.

\begin{figure*}
        \centering
	    \includegraphics[scale=0.4]{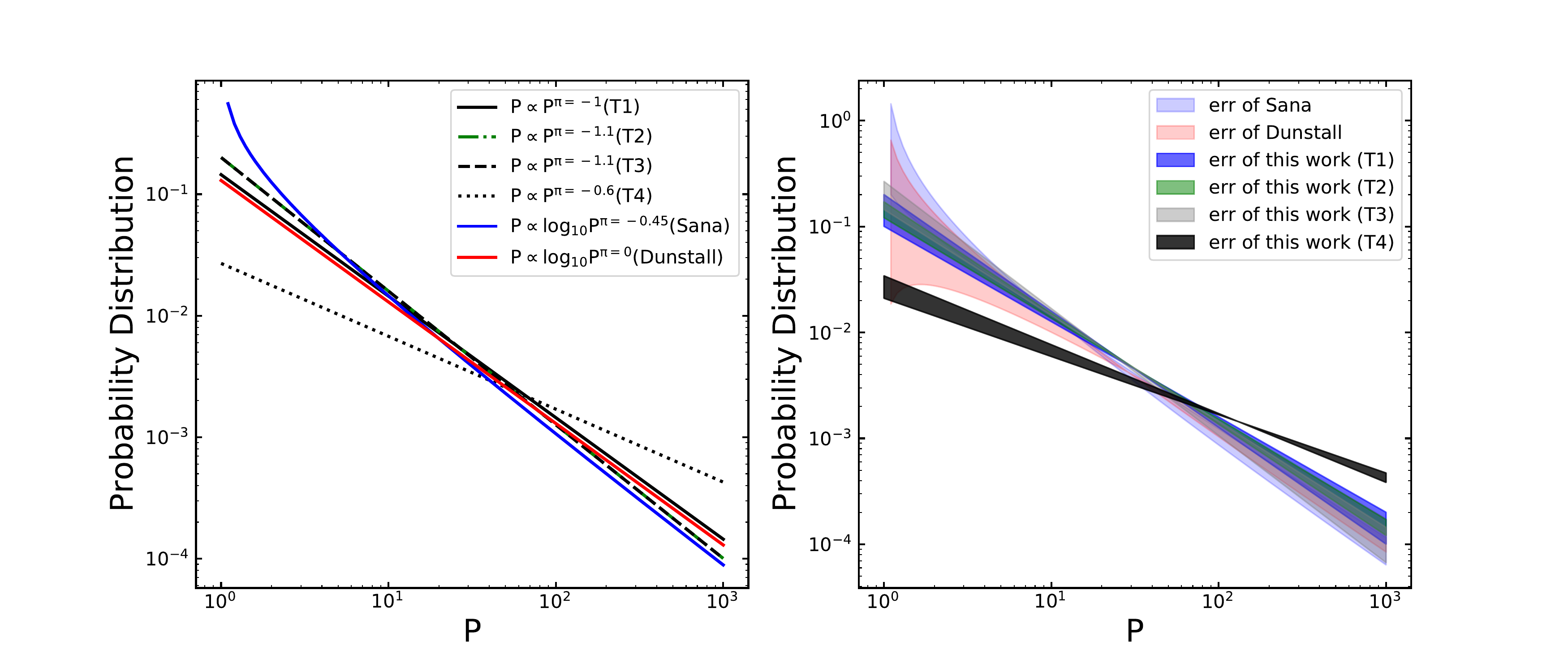}	
		\caption{Comparisons of orbital period distribution with the results from previous works. In the left panel, the blue line represents the result of \cite{2013Sana}, and the red one represents that of \cite{2015Dunstall}. The black and green lines represent the result of this work. The right panel displays the upper and lower uncertainty orbital period distribution for each work.
		}\label{fig:P_result}
\end{figure*}

We are also interested in investigating the distribution of the orbital period as a consequence of spectral types (on $T_\mathrm{eff}$). We then plotted the orbital period distribution in the left panel of Fig.~\ref{fig:P_result} for sample stars in groups T1 to T4 (shown in solid black line, green dashed line, black dashed line, and black dotted line, respectively) as well as similar over-plotted distribution by adopting the results repeated from \cite{2013Sana} in blue and \cite{2015Dunstall} in red. We plotted each work's orbital period distribution with the upper and lower uncertainty ranges by adding and subtracting the uncertainty from the best-estimated values, respectively, in the right panel of Fig.~\ref{fig:P_result}. It seems that the distribution of group T4 is more flat. 

\begin{figure}
        \centering
	    \includegraphics[scale=0.5]{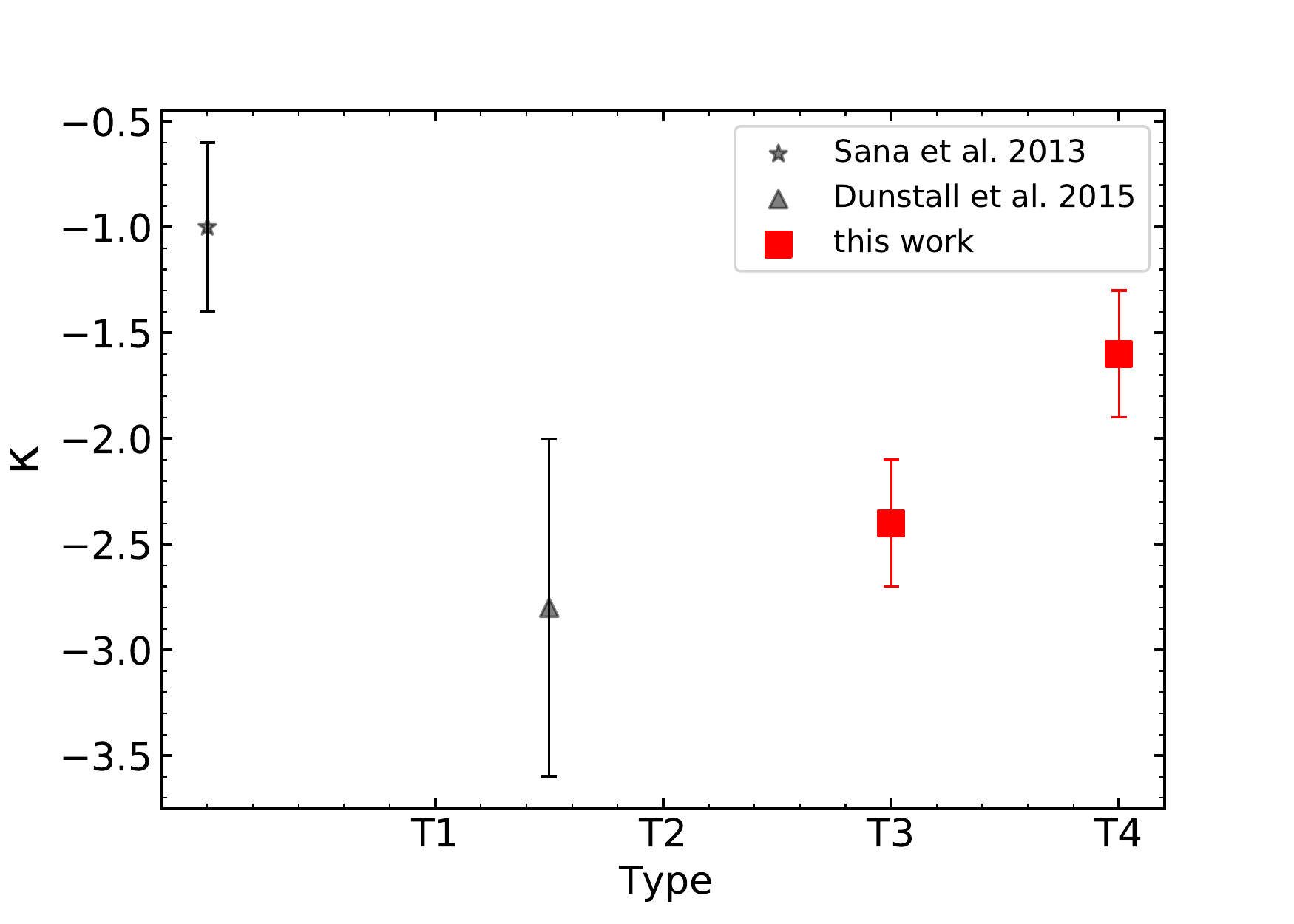}	
		\caption{Comparison of $\kappa$ in T3 and T4 with previous works. The star mark and the triangle represent the result of \cite{2013Sana} and \cite{2015Dunstall}, respectively.}\label{fig:kappa_result}
\end{figure}

Our $\kappa$ (red squares) distributions are shown in Fig.~\ref{fig:kappa_result}. 
The star mark is the result of \cite{2013Sana}, and the triangle represents the result of \cite{2015Dunstall}. All of them have large error bars, and no significant correlation is found. 
 
\section{CONCLUSIONS}\label{sec:conclusions}
We identified 9,382 early-type stars from the LAMOST-MRS dataset based on the EWs. These early-type stars were roughly classified into four groups of T1, T2, T3, and T4 based upon their $T_\mathrm{eff}$ in descending order. The number of stars in each group is 1138, 1092, 2649, and 4503, respectively. Then we calculated the relative radial velocity of multi-epoch early-type stars by using the Maximum Likelihood Estimation method and then identified spectroscopic binaries with significant $\Delta$RV. The observed binary fractions of these four classifications are $24.6\%\pm0.5\%$, $20.8\%\pm0.6\%$, $13.7\%\pm0.3\%$, and $7.4\%\pm0.3\%$, respectively.

We used a Monte-Carlo method to correct observational bias and constrain the intrinsic properties, including the binary fraction, the distributions of orbital period and the mass ratio. The intrinsic binary fractions of stars residing in T1 to T4 are 68$\%\pm8\%$, 52$\%\pm3\%$, 44$\%\pm6\%$, and 44$\%\pm6\%$, respectively. The orbital period distributions follow the power-low of $f(P)\propto$ $P^{\pi}$, and the ${\pi}$ values for each groups are -1$\pm0.1$, -1.1$\pm0.05$, -1.1$\pm0.1$, and -0.6$\pm0.05$, respectively. The mass ratio distributions follow the power-low of $f(q) \propto q^\kappa$, the $\kappa$ is unconstrained for group T1 and T2 while for T3 and T4 we reported the $\kappa$ has values of -2.4$\pm0.3$ and -1.6$\pm0.3$, respectively.

Based on the binary fraction results, the early-type massive stars are likely to have a higher chance of hosting companion stars compared to late-type, low-mass stars.
The multiplicity properties as a basic physical input of binary population synthesis not only have an impact on the final results of binary population synthesis but also allow us to better understand the evolution of massive stars.

For future work, we expect to better constrain these multiplicities parameters as more observations will be made possible.

\begin{acknowledgements}
This work is supported by the Natural Science Foundation of China (Nos.\ 11733008, 12090040, 12090043, 11521303, 12125303), by Yunnan province, by the National Ten-thousand talents program. C.L.\ acknowledges National Key R$\&$D Program of China No.\ 2019YFA0405500 and the NSFC with grant No.\ 11835057.

The authors gratefully acknowledge the ``PHOENIX Supercomputing Platform" jointly operated by the Binary Population Synthesis Group and the Stellar Astrophysics Group at Yunnan Observatories, Chinese Academy of Sciences.

Guoshoujing Telescope (the Large Sky Area Multi-Object Fiber Spectroscopic Telescope LAMOST) is a National Major Scientific Project built by the Chinese Academy of Sciences. Funding for the project has been provided by the National Development and Reform Commission. LAMOST is operated and managed by the National Astronomical Observatories, Chinese Academy of Sciences.

This work is also supported by the Key Research Program of Frontier Sciences, CAS, Grant No.\ QYZDY-SSW-SLH007 and the science research grants from the China Manned Space Project with No.\ CMS-CSST-2021-A10

\end{acknowledgements}

\bibliographystyle{raa} 
\bibliography{msRAA-2019-0285}

\end{document}